\documentclass{aa}  
\bibpunct{(}{)}{;}{a}{}{,} 
\usepackage{graphicx}
\usepackage{txfonts}
\usepackage{hyperref}
\newcommand{\kp}{$K_\mathrm{P}$}
\newcommand{\kstar}{$K_\mathrm{S}$}

\newcommand{\mjup}{$M_\mathrm{Jup}$}
\newcommand{\rjup}{$R_\mathrm{Jup}$}
\newcommand{\mstar}{$M_\mathrm{S}$}
\newcommand{\msun}{$M_\sun$}
\newcommand{\mplan}{$M_\mathrm{P}$}
\newcommand{\kms}{km~s$^{-1}$}
\newcommand{\micron}{$\mu$m}
\newcommand{\water}{H$_2$O}
\newcommand{\methane}{CH$_4$}
\newcommand{\lapserate}{d$T$/d$\log_{10}(p)$}
\newcommand{\mpvalue}{$M_\mathrm{P}=(0.98\pm0.04)~M_\mathrm{Jup}$}
\newcommand{\kpvalue}{$K_\mathrm{P}=(142.8\pm3.4)$ km~s$^{-1}$}
\newcommand{\mratio}{$(1268\pm36)$}
\newcommand{\ivalue}{$(67.7\pm4.3)$ degrees}
\newcommand{\vovalue}{$V_\mathrm{orb}=(154.4 \pm 1.7)$ km~s$^{-1}$} 

\begin{document} 

\title{Carbon monoxide and water vapor in the atmosphere \\ of the non-transiting exoplanet HD~179949~b\thanks{Based on observations collected at the ESO Very Large Telescope during the Program 186.C-0289}}
\author{M. Brogi\inst{1} \and R. J. de Kok\inst{1,2} \and J. L. Birkby\inst{1} \and H. Schwarz\inst{1} \and I. A. G. Snellen\inst{1}}
\institute{Leiden Observatory, Leiden University, P.O. Box 9513, 2300 RA Leiden, The Netherlands\\\email{brogi@strw.leidenuniv.nl}
\and
SRON Netherlands Institute for Space Research, Sorbonnelaan 2, 3584 CA Utrecht, The Netherlands\\}

\date{Received / Accepted}

\abstract
   {In recent years, ground-based high-resolution spectroscopy has become a powerful tool for investigating exoplanet atmospheres. It allows the robust identification of molecular species, and it can be applied to both transiting and non-transiting planets. Radial-velocity measurements of the star HD~179949 indicate the presence of a giant planet companion in a close-in orbit. The system is bright enough to be an ideal target for near-infrared, high-resolution spectroscopy.}
   {Here we present the analysis of spectra of the system at 2.3~\micron, obtained at a resolution of $R~\sim100,000$, during three nights of observations with CRIRES at the VLT. We targeted the system while the exoplanet was near superior conjunction, aiming to detect the planet's thermal spectrum and the radial component of its orbital velocity.}
   {Unlike the telluric signal, the planet signal is subject to a changing Doppler shift during the observations. This is due to the changing radial component of the planet orbital velocity, which is on the order of 100-150 \kms\ for these hot Jupiters. We can therefore effectively remove the telluric absorption while preserving the planet signal, which is then extracted from the data by cross correlation with a range of model spectra for the planet atmosphere.}
   {We detect molecular absorption from carbon monoxide and water vapor with a combined signal-to-noise ratio (S/N) of 6.3, at a projected planet orbital velocity of \kpvalue, which translates into a planet mass of $M_\mathrm{P}=(0.98\pm0.04)$ Jupiter masses, and an orbital inclination of $i=$~\ivalue, using the known stellar radial velocity and stellar mass. The detection of absorption features rather than emission means that, despite being highly irradiated, HD~179949~b does not have an atmospheric temperature inversion in the probed range of pressures and temperatures. Since the host star is active ($R^\prime_{HK} > -4.9$), this is in line with the hypothesis that stellar activity damps the onset of thermal inversion layers owing to UV flux photo-dissociating high-altitude, optical absorbers. Finally, our analysis favors an oxygen-rich atmosphere for HD~179949~b, although a carbon-rich planet cannot be statistically ruled out based on these data alone.}
   {}


\titlerunning{CO and \water\ in the atmosphere of HD 179949 b}
\authorrunning{Brogi et al.}

\maketitle

\section{Introduction}
Since the mid-1990s, the exoplanet community started to use ground-based high-resolution spectroscopy as a tool for measuring signals from exoplanet atmospheres \citep{ccam99, charb99}. It allows molecular bands to be resolved into the individual lines, offering a robust way of identifying molecular species by line matching. It is also able to detect the planet orbital motion in the form of a Doppler-shifted spectrum. This means that the radial velocity of the planet can be directly measured, and the star + planet system can be treated in the same way as a spectroscopic binary. While the first observations of exoplanets at high spectral resolution were focused on detecting reflected starlight from hot Jupiters, it soon became clear that most of these planets might have a very low albedo \citep{mar99, sud00}. Similar searches for thermal emission \citep[e.g.,][]{wie01} were conducted at a lower resolving power ($R<50,000$) and were limited by the narrow spectral range of near-infrared instruments. Therefore, these early studies were only able to put upper limits on the detected planet signal, and ground-based high-resolution spectroscopy remained unsuccessful at detecting signatures from an exoplanet atmosphere for more than a decade.

Transiting planets can be observed at high spectral resolution while crossing the stellar disk (transmission spectroscopy), when a small fraction of the stellar light filters through their atmospheres, but also just before and after secondary eclipse, when the illuminated planet hemisphere is facing the observer (dayside spectroscopy). In the latter case, the thermal emission from the planet is targeted directly, meaning that dayside spectroscopy can be applied to non-transiting planets as well. Direct imaging and direct spectroscopy, which have recently achieved great successes in characterizing non-transiting planets \citep{kon13}, can so far be applied only to young, self-luminous planets on large orbits. High-resolution spectroscopy offers a complementary method capable of studying the atmospheres of close-in, evolved non-transiting planets, and of determining their orbital inclination and true mass. 

By using the CRyogenic Infra-Red Echelle Spectrograph \citep[CRIRES,][]{crires} at the ESO Very Large Telescope (VLT) facility, the first molecular detections at high spectral resolution were obtained in recent years. By observing a transit of HD 209458 b at 2.3~\micron, the radial velocity of the exoplanet was measured \citep{sne10} by tracing the Doppler-shifted carbon monoxide absorption while the planet was transiting the host star. Dayside spectroscopy in the same wavelength range revealed molecular absorption in the atmosphere of the non-transiting exoplanet $\tau$ Bo\"otis b \citep{bro12,rod12}, whose mass and orbital inclination were also determined by means of the measured planet radial velocity. Molecular absorption from CO and possibly water was also observed in the dayside spectrum of 51 Peg b \citep{bro13}, although one of the three nights of observations does not show a signal and therefore more data are required in order to confirm the detection. Finally, the dayside spectrum of the transiting planet HD 189733 b was also investigated at high-resolution around 2.3~\micron\ and 3.2~\micron. The $K$-band data revealed CO absorption \citep{kok13}, and the $L$-band data resulted in the detection of water vapor absorption \citep{bir13}, which demonstrates the robustness of high-resolution spectroscopy even when searching for more complicated molecules such as H$_2$O and in spectral regions heavily contaminated by telluric absorption.

Here we present high-resolution dayside spectroscopy of the non-transiting planet HD 179949 b. We detect carbon monoxide and water vapor in the planet atmosphere, and solve for the planet mass and system inclination. The system HD 179949 is introduced in Sect.~\ref{sys-description}, while Sect.~\ref{obs-data} presents our observations and data analysis. Sect.~\ref{ccorr} explains how we cross correlate the data with model spectra in order to extract the planet signal, and the results of our analysis are presented in Sect.~\ref{results}. The interpretation of our measurements and future prospects of high-resolution spectroscopy for exoplanet studies are presented in Sect.~\ref{discussion}.

\section{The system HD 179949}\label{sys-description}

The F8 main-sequence star HD~179949 has a mass of \mstar~= (1.181$^{+0.039}_{-0.026}$)~\msun\ and a radius of $R_\mathrm{S}$~= (1.23~$\pm$~0.04)~$R_\odot$ (Takeda et al. 2007). Interestingly for these observations, the relatively high stellar effective temperature ($T_\mathrm{eff}$~= 6168~$\pm$~44~K) and projected rotational velocity (7.0~$\pm$~0.5~\kms) ensure that no significant stellar absorption is present in the wavelength range covered by our data, especially CO absorption which would cause a spurious cross-correlation signal when the planet and the star have a similar radial velocity \citep[i.e., around superior conjunction, see][]{bro13}. This star is known to host a planet from radial velocity (RV) measurements (Tinney et al. 2001). Since no transit of HD~179949~b is detected, only a lower limit of $m\sin(i)$~= (0.902~$\pm$~0.033) Jupiter masses ($M_\mathrm{Jup}$) can be put on its mass, with the orbital inclination $i$ being unknown.

\citet{cow07} observed the phase curve of HD~179949 with Spitzer/IRAC at 3.6~\micron\ and 8.0~\micron. At the longer wavelength, they detected peak-to-peak flux variations of $(1.41\pm0.33)\times10^{-3}$ relative to the average flux from the system, in phase with the orbit of the planet. By assuming a planet radius (1 $< R_\mathrm{P} < 1.2)~R_\mathrm{Jup}$ and a Bond albedo close to zero, they were able to constrain the recirculation efficiency as a function of system inclination, and concluded that less than 21\% of the incident energy must recirculate to the night side.

\citet{bar08} used CRIRES at a resolution of $R$~$\sim$~50,000 to observe HD~179949 during two nights in the wavelength range 2.1215-2.1740~\micron. After testing for molecular absorption and emission, they were able to rule out absorption features at a 99 per cent confidence level, for a planet-to-star contrast ratio of $\sim3\times10^{-4}$. As shown in Sect.~\ref{results}, our observations detect molecular absorption from HD~179949~b, at a contrast ratio of roughly $\sim10^{-4}$, which is consistent with the above upper limit. 

Finally, the star HD~179949 is an active star, with an activity index of $\log(R^\prime_{HK})$~=~$-4.720\pm0.031$ \citep[][supplementary information]{wats10}. By analyzing the variability of the \ion{Ca}{ii} K and H emission, \citet{shk03} reported evidence for the activity of HD~179949 being in phase with the planet orbit. \citet{fares12} studied the large-scale magnetic field of the star, and concluded that the chromospheric activity of HD~179949 is mainly modulated by the stellar rotation, although there is some evidence of an additional modulation with the period of the planet companion. A similar conclusion was obtained by \citet{gurdemir12}. These results support the hypothesis that HD~179949 experiences planet-induced emission enhancement, as a result of the strong planet-star interaction.

\section{Observations and data analysis}\label{obs-data}

\subsection{Telescope, instrument and observations}\label{obs}

As part of the Large Program 186.C-0289, 1000 spectra of the system HD 179979 (K~=~4.94 mag) were obtained during the nights of July 17, 2011 (398 spectra, 5.5 hours of observations excluding overheads), July 19, 2011 (388 spectra, 5.4 hours of observations), and August 22, 2011 (214 spectra, 3 hours of observations) with CRIRES, mounted at the Nasmyth-A focus of the VLT UT1 (Antu).

The spectrograph was set to cover the wavelength range 2287.5-2345.4 nm, roughly centered on the 2-0 R-branch of carbon monoxide. Possible additional molecular absorption from water vapor and methane is also present at these wavelengths. We employed CRIRES at the maximum resolution ($R=100,000$) by using the 0.2$^{\prime\prime}$ slit. In order to maximize the instrument throughput, the Multi Application Curvature Adaptive Optic system \citep[MACAO,][]{macao} was used. The observations were conducted by nodding the spectrograph 10\arcsec\ along the slit, in a standard ABBA pattern, for accurate background subtraction.

Unlike our previous observations of $\tau$ Bo\"otis and 51 Pegasi \citep{bro12, bro13}, the observing conditions differed considerably between the three nights for this target. Firstly, HD~179949 reaches almost the zenith from Paranal, and therefore the airmass varies considerably during a full night of observations, which causes spectra taken at airmass $\sim$2 to have approximately half the signal-to-noise ratio (S/N) of those taken around culmination. Moreover, during the night of August 22, 2011, HD 179979 was observed as a backup target because of pointing restrictions, resulting in non-optimal performances of the MACAO AO system, and a poorer S/N. Finally, during the first two nights, the spectra were taken in groups of 100. At the end of each group, the target was reacquired by the telescope and the AO loop temporarily opened, resulting in small temporal gaps, reducing the S/N of the first spectra of each series, and impacting the overall instrument stability. These factors are taken into account during the data analysis, as explained in Sect.~\ref{align}.

\subsection{Extraction of the one-dimensional spectra}

The basic calibration and extraction of the spectra was performed with the CRIRES pipeline version 2.1.3, in combination with both the GUI Gasgano (version 2.4.0) and the command-line interface esorex (version 3.9.0). 

Standard calibration frames taken in the morning after each night of observations were used for flat-fielding and dark correction. The non-linearity measurements available in the CRIRES archive were also used.
Each frame was dark-subtracted, non-linearity corrected and flat-fielded. Then each couple of nodded observations (AB or BA) was combined into a single spectrum for accurate background subtraction. The one-dimensional spectra were finally obtained by optimal extraction \citep{hor86}, resulting in 199, 194, and 107 spectra respectively for the three nights of observations. These were subsequently analyzed independently in order to account for different atmospheric and instrumental conditions. 

\subsection{Bad-pixel correction and wavelength calibration}\label{align}

\begin{figure}
\centering
\includegraphics[width=0.5\textwidth]{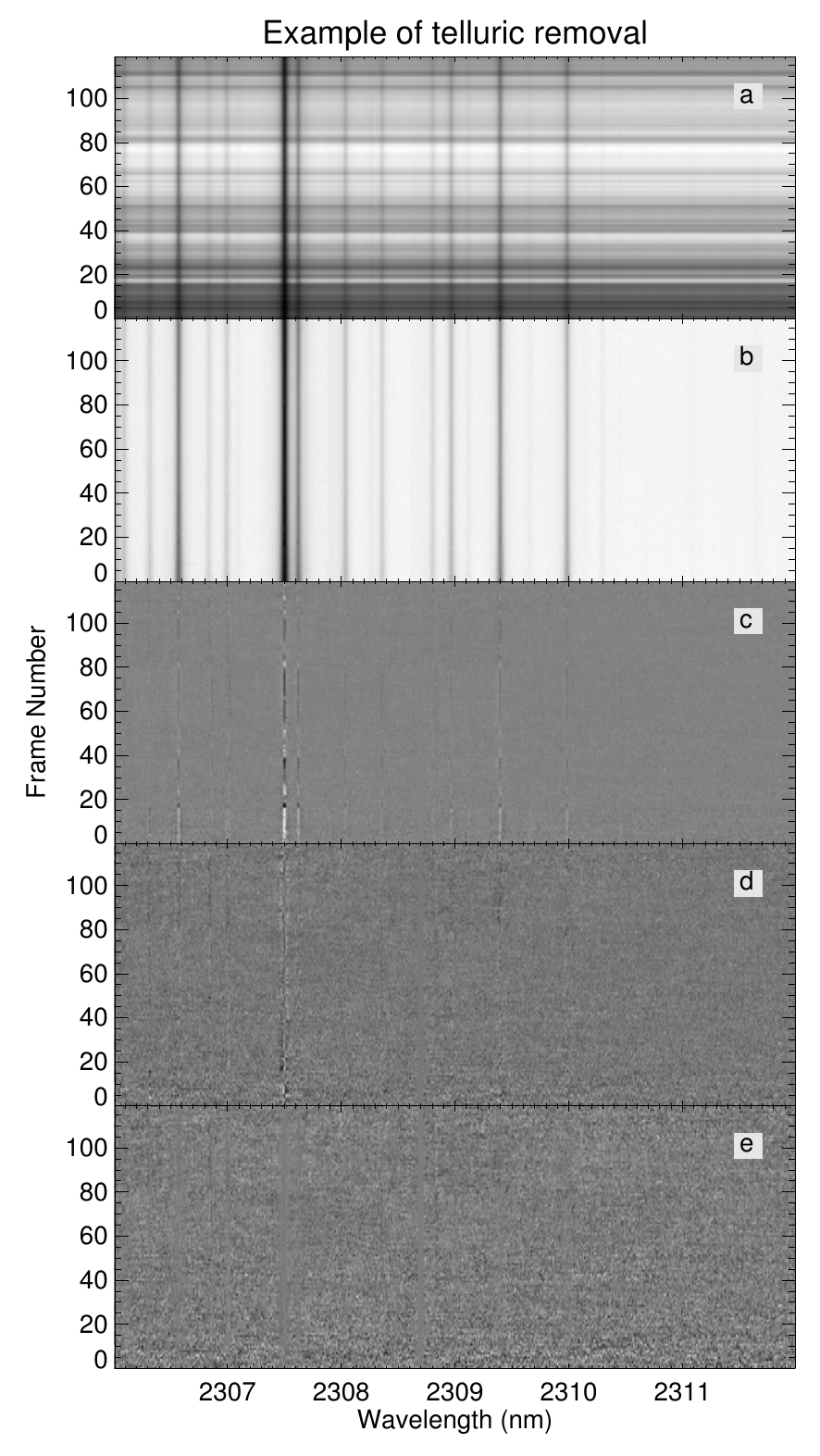}
\caption{Example of telluric lines removal. Panel (a) shows a sequence of 120 spectra of HD 179949 observed during the night of July 19, 2011. A small portion of detector 2 of CRIRES is shown, corresponding to the wavelength range 2306-2312 nm. The spectra are first normalized by their median value (b). Subsequently, the flux in each column is divided through its linear fit with airmass (c). A linear combination of the strongest residuals is utilized to detrend second-order effects (d) and finally each column is normalized by its standard deviation squared (e).}
\label{tell-rem}
\end{figure}

Subsequently, the data analysis was performed by employing our own dedicated set of routines written in IDL. As for all our previous high-resolution data, the spectra of each night of observation and each of the CRIRES detectors were organized in matrices with pixel number (wavelength) on the horizontal axis and frame number (time) on the vertical axis, of which an example is shown in Fig.~\ref{tell-rem} (panel $a$). 

The first step in the data analysis is to correct for detector cosmetics and cosmic rays. Isolated bad-pixels were replaced by their spline-interpolated value across the neighboring pixels in the spectral direction. Small groups of adjacent bad-pixels (typically 2 or 3 pixels along the spectral axis) were corrected by linear interpolation. Finally, there was not enough information to correct for large groups (> 5) of adjacent bad pixels. These were therefore masked in all the subsequent steps of the data analysis.

The spectra were then aligned to a common wavelength scale, which is in this case that of the spectrum with the highest S/N in the series. We fit the centroid position of the deepest lines in each spectrum, and computed the difference with those of the reference spectrum. This difference was then fit with a linear function of pixel number and the fit was utilized to shift each spectrum via spline interpolation.
Under standard observing conditions, the wavelength solution of CRIRES spectra varies smoothly as a function of time, and the resulting shifts are generally at the sub-pixel level. For these observations, because of the re-acquisition of the telescope (see Sect.~\ref{obs} above), the wavelength solution shifted by 1-2 pixels each time a series of 100 spectra was taken. We therefore discarded 2 pixels at the edges of the spectra, in order to prevent artifacts from the interpolation to the new pixel scale. 
Finally, with the spectra all aligned to the same reference, we determined the common wavelength solution by comparing the pixel position of the telluric lines in the spectra with their vacuum wavelengths listed in the HITRAN database. We find that a quadratic fit (pixel, wavelength) is sufficient for this purpose, with typical residuals having r.m.s. of 0.15-0.30~\kms\ (or 0.1-0.2 pixels). The wavelength solution was determined for each of the CRIRES detectors separately. We note that detectors 1 and 4 suffer from the so-called ``odd-even effect'', which is a non-linear change in gain between odd and even columns along the spectral direction\footnote{\url{http://www.eso.org/sci/facilities/paranal/instruments/crires/doc/VLT-MAN-ESO-14500-3486_v93.pdf}}. Although for most of the science uses of CRIRES this effect can be corrected by using non-linearity measurements, in our case detector 4 still showed residual odd-even signatures after being calibrated. These residuals would significantly affect the data when the spectra are aligned to the same reference frame, especially for these observations in which the target is reacquired multiple times. We therefore discarded the fourth detector for the remainder of the analysis.

\subsection{Removal of telluric lines}\label{rm_tell_std}

In these ground-based, high-resolution observations, telluric lines dominate over the planet signal in the data by several orders of magnitude. Unlike the planet signal, telluric lines are stationary in wavelength and can therefore be removed (i.e.\ normalized) by modeling the flux of each resolution element (i.e.\ each column of the matrix) in time. In previous work \citep{kok13,bir13}, we have tested several ways of performing this task, including Singular-Value Decomposition and the \textsc{Sysrem} algorithm \citep{tam05}. Here we followed the approach of \citet{bro12} and detrended the data for airmass. This approach works well when the contribution from stellar lines is negligible and the dominant factor driving the instrument performances (resolution, stability, and throughput) is the airmass.

The various stages of the telluric removal are shown in Fig.~\ref{tell-rem}. We first normalized each spectrum in the series by its median value, in order to correct for varying throughput (panel $b$). We then fit the logarithm of the flux in each of the detector pixels (i.e., each column in the matrix) with a linear function of airmass. This step removed the main airmass dependence, but left strong residuals in the data, at the level of 1-2 percent, at the position of some of the telluric lines (panel $c$). These second order effects are due to differential airmass effects (e.g., changes in the atmospheric water vapor content) and/or changes in the instrument profile during the night. In order to remove these effects, we sampled the strongest residuals at the position of 2-3 telluric lines per detector, and we modeled every other pixel as a linear combination of these, determined through linear regression.
This very effectively detrended most of the telluric contamination in the data (panel $d$). Away from the cores of deep absorption lines, we achieve an r.m.s. which is at most $\sim$15\% higher than the photon noise. At the core of strong absorption lines, the r.m.s. is a factor of 2-4$\times$ higher than the photon noise, but these columns account for a small portion of the data, so that on average we obtain noise levels within 20\% from the photon noise.
In order to account for the different signal-to-noise as a function of wavelength, we normalized each column in the data matrix by its standard deviation squared (panel $e$). 

\section{Extraction of the planet signal}\label{ccorr}

Even after applying the telluric removal explained in Sect.~\ref{rm_tell_std}, the planet signal is still buried in the noise, as shown in the lowest panel of Fig.~\ref{tell-rem}. Depending on the molecular species contributing to the opacity at 2.3~\micron\ and their relative abundances, the planet spectrum consists of tens to hundreds of molecular lines Doppler-shifted with respect to their rest-frame frequencies according to the radial component of the planet orbital velocity. In order to enhance the planet signal, we funneled the information of these many molecular lines toward a single function, which was achieved by cross-correlating the data with model spectra for the planet atmosphere. The rest of this Section explains how these models and the cross correlation were computed. 

\subsection{Model spectra}\label{models}

\begin{figure*}[htb]
\centering
\includegraphics{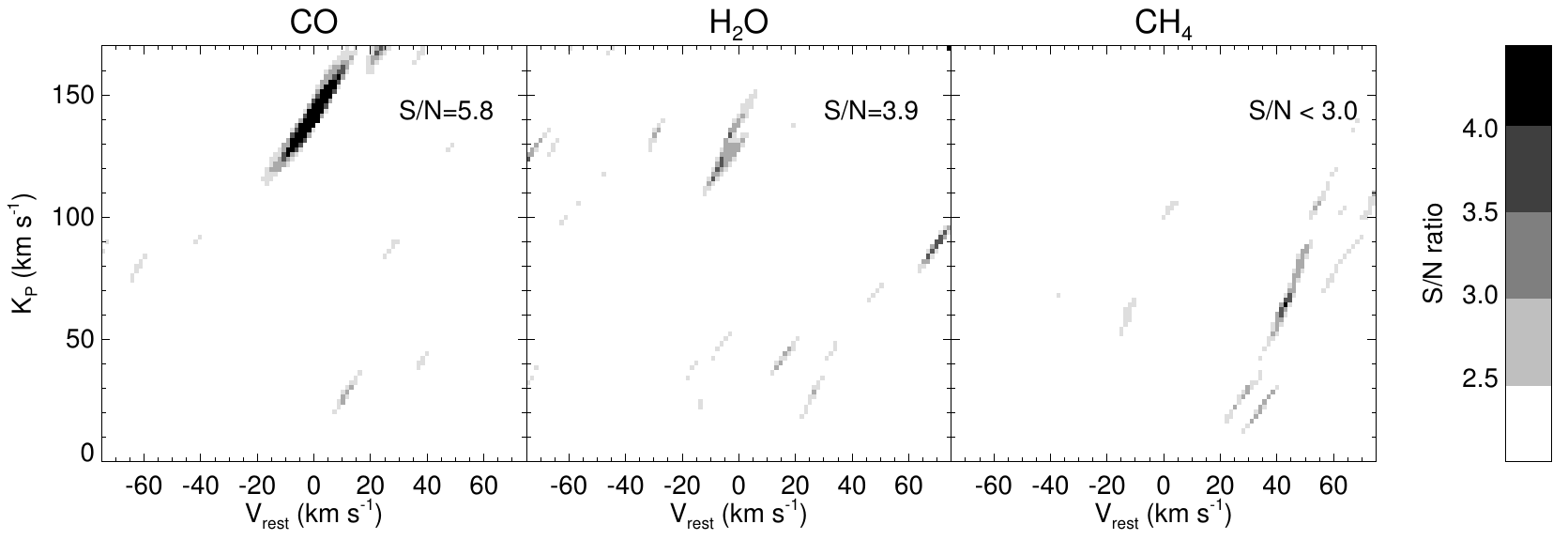}
\caption{Total strength of the cross-correlation signal, as function of rest-frame velocity $V_\mathrm{rest}$ and planet projected orbital velocity \kp, for atmospheric models containing CO only (left panel), H$_2$O only (mid panel), and CH$_4$ only (right panel). It shows that carbon monoxide and water vapor are detected in the atmosphere of HD~179949~b around 2.3~\micron, while methane is not. The signal-to-noise is computed by dividing the value of the cross correlation at each of the positions in the diagram by the standard deviation of the cross-correlation noise. Positive and negative cross-correlation signals are both plotted.}
\label{single_mols}
\end{figure*}

Model spectra of HD 179949 b were constructed using an average temperature/pressure ($T$/$p$) profile describing the vertical structure of the visible hemisphere of the planet atmosphere. This profile was parametrized by two points in the ($T$, $p$) space, namely ($T_1$, $p_1$) and ($T_2$, $p_2$). The atmosphere was assumed to be isothermal for pressures $p > p_1$ and $p < p_2$, at temperatures $T_1$ and $T_2$ respectively. For $p_1 > p > p_2$, a constant lapse rate was adopted, given by
\begin{equation}
\frac{\mathrm{d}T}{\mathrm{d}\log_{10}(p)} = \frac{T_1 - T_2}{\log_{10}(p_1)-\log_{10}(p_2)}.
\end{equation}
Inverted and non-inverted $T$/$p$ profiles were therefore obtained by choosing $T_2 > T_1$ and $T_2 < T_1$, respectively. The model spectra are the result of line-by-line calculations including H$_2$-H$_2$ collision-induced absorption \citep{bory01, bory02} and absorption from three trace gases, CO, H$_2$O, and CH$_4$, which are the dominant sources of opacities at these wavelengths \citep{madhu12, moses13}. Line data for the trace gases were taken from HITEMP 2010 for H$_2$O and CO \citep{roth10}, and HITRAN 2008 for CH$_4$ \citep{roth09}. A Voigt line profile was used for the calculations, with broadening coefficients obtained from the HITEMP/HITRAN databases and accounting for pressure broadening only. As a result of the high temperatures involved, this carries the dominant contribution to the line profiles. The methane line list is known to contain two strong but apparently spurious absorption lines in this wavelength range (S.~Yurchenko and J.~Tennyson, private communication), at 2326.90 nm and 2335.14 nm. These lines were therefore removed from the input of our models. We also compared the strengths of the molecular lines in the HITRAN and EXOMOL databases, finding a good agreement between them. Because of computational reasons and to the degeneracies affecting these high-dispersion observations (see Sect.~\ref{degeneracies} below), we did not explore a full grid of parameters by varying temperatures, pressures, and molecular abundances at the same time. We instead tested pressures in the range $(-4.5 < \log_{10}(p_2) < -1.5)$ in steps of 1.0 dex, while $p_1$ was fixed to 1 bar, which is approximately the expected continuum level for these gaseous planets \citep{remco14}. 

We first computed planet spectra by including a single trace gas. We utilized the upper limit of $\sim$20\% on the recirculation efficiency set by \citet{cow07} in combination with Eq.~4 of \citet{cow11} in order to get an estimate of the dayside temperature of HD 179949 b, which was found to be $T_\mathrm{eq} \simeq$ 1950 K. 
Different assumptions for the recirculation efficiency would lead to dayside temperatures between 1570 K (full recirculation) and 2000 K (no recirculation).
We set $T_1 = T_\mathrm{eq}$, and explored $T_2 = [1450, 1800, 2150]$ K. The chosen Volume Mixing Ratios (VMRs) of the three molecules were VMR = [10$^{-6}$, 10$^{-5}$, 10$^{-4}$]. This first set of models allowed us to determine which species are significantly detected, and to determine the presence or the absence of an inversion layer, as discussed in Sect.~\ref{res-single}. We then computed an additional grid of models with the same choice of $p_1$, $p_2$, and $T_1$, but with only non-inverted profiles ($T_2 = 1450$ K), since we did not detect any significant cross-correlation from models with $T_2 > T_1$. In this case, the VMR of CO was set to 10$^{-4.5}$, which roughly matches the measured line contrast ratio of the planet spectrum (see Sects.~\ref{degeneracies} and \ref{res-single}). The abundances of H$_2$O were varied in the range ($-8.5 < \log_{10}$(VMR) $< -2.5$) and those for CH$_4$ in the range ($-9.5 <\log_{10}$(VMR) $< -4.5$), with steps of 1 dex. This choice covers the whole range of relative abundances for a hot Jupiter with the above $T_\mathrm{eq}$, as a function of its C/O ratio \citep{madhu12}. As explained in Sect.~\ref{res-three}, we determined the best-fitting model for the planet atmosphere, the planet mass, the orbital inclination, and the atmospheric C/O ratio by using this second grid of models.

\subsection{Cross correlation and signal retrieval}
We cross correlated each model spectra with the telluric-removed data by utilizing a lag vector corresponding to radial velocities (RVs) in the range $-250<V<250$~\kms, in steps of 1.5~\kms. This range was chosen in order to cover all the possible RVs for the planet, with the step size approximately matching the pixel scale of the instrument detectors at this wavelength.

For each RV value, we Doppler-shifted the model spectrum and computed its cross correlation with the time series of the observed spectra. In this way we obtained the strength of the cross-correlation signal as a function of radial velocity and time, that is CCF($V, t$). This computation was first performed for each night and each detector separately. Subsequently, the CCF matrices of the three CRIRES detectors were summed after weighting each CCF by the S/N of the corresponding spectrum. As we explained above, the fourth CRIRES detector was excluded from the analysis.
Finally, the CCFs of the three nights were summed in time. This last step was performed by first shifting each CCF to the rest frame of the planet, which implies the knowledge of the planet radial velocity curve, the velocity of the observer with respect to the barycenter of the solar system ($V_\mathrm{obs}(t)$) and the systemic velocity of HD~179949, that is $V_\mathrm{sys} = (-24.35 \pm 0.18)$~\kms \citep{but06}:
\begin{equation}\label{pl_orb_vel}
V_\mathrm{P}(t) = K_\mathrm{P}\sin[2\pi\varphi(t)] + V_\mathrm{obs}(t) + V_\mathrm{sys}.
\end{equation}
In the above Equation $\varphi$ is the planet orbital phase, and \kp\ is the amplitude of the planet orbital radial velocity. Since \kp\ is unknown for a non-transiting planet, we assumed a range of (0~$\le$~\kp~$\le$~170)~\kms, which corresponds to all the possible orbital inclinations of the system, plus some unphysical values ($\sin~i>1$) as a sanity check. In the case of our best-fitting model (see Sect.~\ref{res-three}), we extended the range to negative \kp. This is done to further assess the robustness of the total cross-correlation signal, as discussed in Sect~\ref{hd179_stats}. For each value of \kp, the planet radial velocity curve was computed, each CCF was shifted to the planet rest frame, and the signal was summed in phase. In this way we obtained the total strength of the cross correlation signal as a function of the planet rest-frame velocity ($V_\mathrm{rest}$) and projected orbital velocity (\kp), which is shown in Figs.~\ref{single_mols} and \ref{all-mols}.
We note that, in the case of HD~179949~b, the exact orbital phase is an additional source of uncertainty in computing Eq.~\ref{pl_orb_vel}. The orbital phase is obtained by taking the fractional part of
\begin{equation}
\varphi(t) = \frac{t - T_0}{P},
\end{equation}
where $P$ is the orbital period and $T_0$ the time of inferior conjunction. From the most accurate orbital solution for the planet \citep{but06} we have $P=3.092514(32)$~days and $T_0=(2,451,001.516\pm0.020)$~HJD, which translates into a 1-$\sigma$ uncertainty of $\Delta\varphi=0.012$ in the orbital phase at the time of observation. Since the systemic and barycentric velocities are known with a much better precision, we iteratively solved for $\varphi(t)$ by first computing the total cross-correlation signal as a function of \kp, and then by applying a shift $\Delta\varphi$ until the planet signal peaked at zero planet rest-frame velocity. This also allowed us to refine the value of $T_0$. In all these computations, we made the assumption that the orbit of HD~179949~b is circular. \citet{but06} presents weak evidence for a low eccentricity of $e=(0.022\pm0.015)$, but we did not measure any significant change in the strength of the planet signal by assuming this solution. Therefore, we adopted $e=0$ through the remainder of the analysis.

\subsection{Degeneracies in the atmospheric models}\label{degeneracies}

In these high-dispersion observations, each spectrum is renormalized for accurate telluric removal, as described in Sect.~\ref{rm_tell_std}. This means that information on absolute fluxes, as well as on broad-band flux variations, is lost. On the other hand, information on the narrow component of the planet spectrum, e.g., the flux ratio between the planet continuum and the core of the lines, is preserved. 

\begin{figure*}[!htb]
\centering
\includegraphics{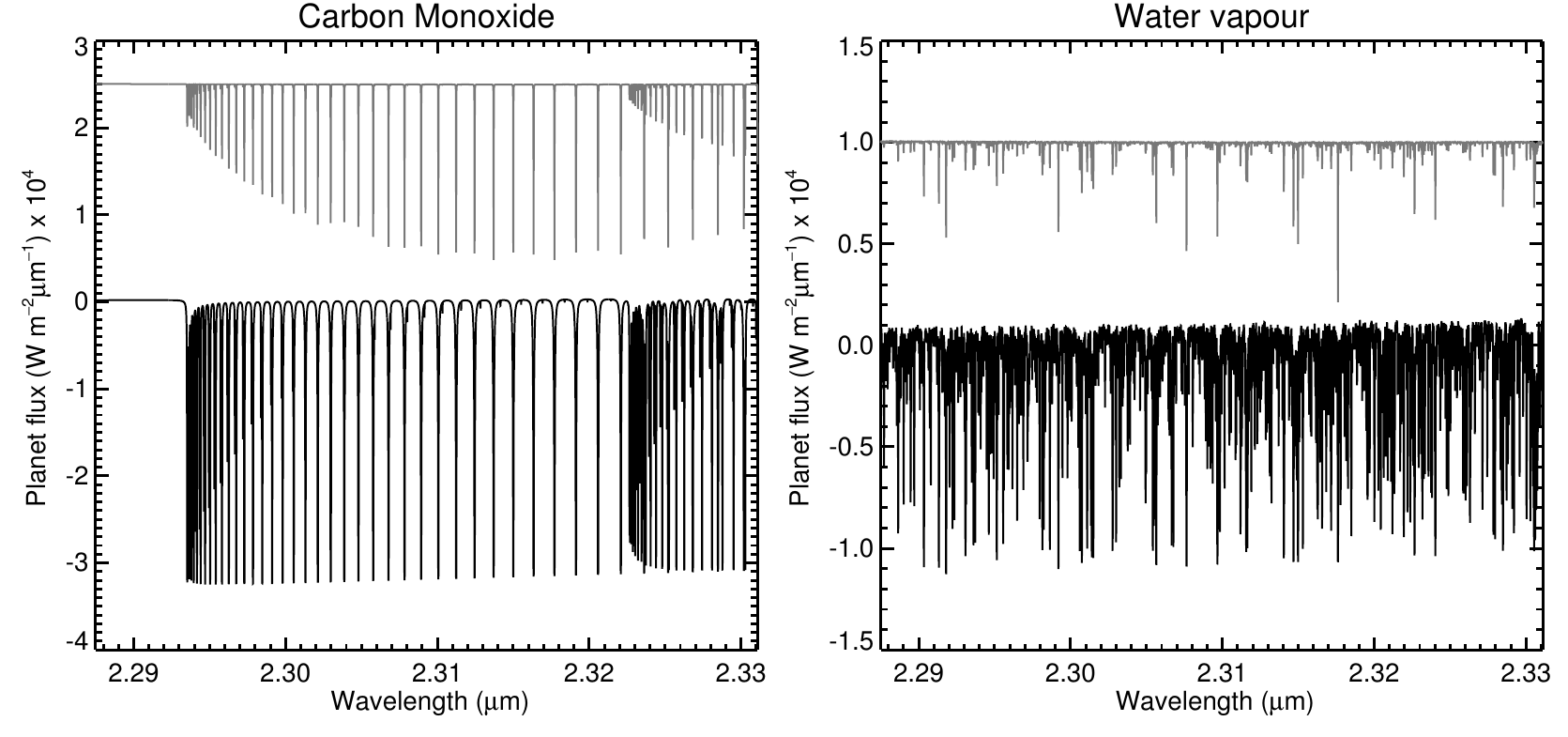}
\caption{Effects of line saturation for models containing pure CO (left panel) and pure H$_2$O (right panel). The models are continuum-subtracted and arbitrarily shifted in flux for visualization purposes. The atmospheric $T/p$ profile is kept constant between the top and bottom models, but the molecular VMR is increased by two orders of magnitude from top to bottom, leading to saturation of the strongest lines. The same effects could be obtained by keeping the abundances constant, and increasing the lapse rate d$T$/d$\log(p)$.}
\label{unsat-sat}
\end{figure*}

By utilizing the cross correlation to extract information from the data, we were able to investigate two main properties of the planet spectrum. Firstly, we measured the relative strength of the molecular lines with respect to each other, which we henceforth refer to as {\it relative line ratios}. This property was estimated directly by cross-correlating the model spectra with the data, and by selecting the model giving the highest total cross-correlation signal. Secondly, we measured the average depth of the planet spectral lines with respect to the stellar continuum flux, which we call {\it average line depth}, in the following way. We scaled the planet model spectrum to units of stellar continuum flux, via
\begin{equation}\label{scaling}
F_\mathrm{scaled}(\lambda) = \frac{\pi F_\mathrm{P}(\lambda)}{F_\mathrm{star}(\lambda)}\left(\frac{R_\mathrm{P}}{R_\mathrm{S}}\right)^2,
\end{equation}
where $F_\mathrm{star}(\lambda)$ is the NextGen low-resolution model spectrum for a HD~179949-like star ($T_\mathrm{eff}=6200$~K, $\log~g=4.5$, solar metallicity), $R_\mathrm{P}$ and $R_\mathrm{S}$ are the planet and stellar radii respectively, and $F_\mathrm{P}(\lambda)$ is the output of our models. The factor of $\pi$ accounts for the integration over the solid angle. The scaled model ($F_\mathrm{scaled}$) was convolved to the CRIRES resolution and subtracted from the data ($F_\mathrm{meas}$) before the removal of telluric lines, via
\begin{equation}\label{scaling}
F_\mathrm{sub}(\lambda) = F_\mathrm{meas}(\lambda)\left[1-F_\mathrm{scaled}(\lambda)\right].
\end{equation}
After performing the full analysis, we then cross-correlated the telluric- and model-subtracted data with the same model. When the average line depth of the model matches the average line depth in the observed spectra, the subtraction is perfect and we obtain zero residual cross-correlation signal. This means that the best-fitting model for the planet atmosphere must give at the same time the highest cross-correlation signal, and zero cross-correlation signal when subtracted from the data.

Even by pairing the study of the relative line ratios and the average line depth as outlined above, these high-resolution observations are affected by degeneracies. The contrast between the planet continuum (deeper in the atmosphere) and the line core (much higher in the atmosphere) reflects a {\sl temperature} difference between the two atmospheric layers. This depends on three quantities: the continuum temperature, the atmospheric lapse rate, and the molecular abundances. Therefore, large degeneracies exist when comparing models to the data. For transiting planets, the degeneracies can be lifted since the continuum temperature can be derived from low-resolution measurements of secondary-eclipse depth. However, this is not possible for non-transiting planets and therefore weaker constraints can be placed through phase curves, as in the case of this planet. 

In addition, $R_\mathrm{P}$ is also unknown for non-transiting planets. Here we assume $R_\mathrm{P}\simeq1.35$~\rjup, which is the mean radius of the known transiting planets with masses in the range 0.9-1.1~\mjup, and orbital periods less than 10 days.

\section{Results}\label{results}

\subsection{Single-molecule detections}\label{res-single}
Fig.~\ref{single_mols} shows the total strength of the cross-correlation signal as a function of the planet rest-frame velocity $V_\mathrm{rest}$ and projected orbital velocity \kp, as obtained from models containing carbon monoxide only, water vapor only, and methane only. CO and H$_2$O are detected, when cross correlating with non-inverted atmospheric models, at a S/N of 5.8 and 3.9 respectively. The level of detection is estimated by dividing the peak value of the cross-correlation by the standard deviation of the noise. No significant cross-correlation is obtained when testing models with thermal inversion. It means that we measure only molecular features in absorption in the atmosphere of HD~179949~b. Finally, models with \methane\ do not result in any cross-correlation signal above the threshold of S/N~=~3. 
The consequences of the measured lack of thermal inversion, and the validity of our models based on a single $T/p$ profile extending to pressures well above typical levels for inversion layers are discussed in Sect.~\ref{discussion}.

We note that the models that contain CO show a different behavior than those based on H$_2$O and CH$_4$ for varying atmospheric vertical structure and molecular abundances. The core of carbon monoxide lines is formed at very high altitudes in the planet atmosphere even for relatively low molecular abundances, resulting in partially-saturated lines due to our assumption of an isothermal planet atmosphere above a certain pressure level. This means that the relative line ratio (defined in Sect.~\ref{degeneracies}) of the CO molecule does not change much with the atmospheric thermal structure or molecular abundances (Fig.~\ref{unsat-sat}, left panel). Conversely, in the case of H$_2$O and CH$_4$, when the temperature decreases steeply with pressure, or for high molecular abundances, the strongest planet spectral lines tend to saturate, therefore altering the relative line ratios (Fig.~\ref{unsat-sat}, right panel). Because CO dominates the cross-correlation signal at 2.3~\micron, the sensitivity of these data to various atmospheric structure is therefore decreased, adding to the degeneracies discussed in Sect.~\ref{degeneracies}. From the analysis of relative line ratios of models containing single molecules, we only observe a marginal preference for models with {\it saturated} line profiles, for both CO and H$_2$O.

When investigating the average line depth as explained in Sect.~\ref{degeneracies}, we find that models with a steep lapse rate (\lapserate~$\ge$~300 K/dex) and VMR(CO) between 10$^{-5}$ and 10$^{-4}$ give zero cross correlation when subtracted from the data. The corresponding line contrast with respect to the stellar continuum is $(1.74\pm0.27)\times10^{-4}$ and $(0.86\pm0.24)\times10^{-4}$, for CO and \water\ respectively. For the analysis described in Sect.~\ref{res-three}, we fix VMR(CO) to the intermediate value of 10$^{-4.5}$.

\subsection{Planet mass and orbital inclination}\label{res-three}
From the previous single-molecule detections, we deduce that the majority of the deepest lines in the dayside spectrum of HD~179949~b comes from CO absorption, which also gives the dominant cross-correlation signal. Additional signal comes from shallower (but more numerous) H$_2$O molecular lines. The contribution from methane appears to be negligible, or below the threshold of detectability, at these wavelengths. As explained in Sect.~\ref{models}, we repeated the cross-correlation analysis by utilizing a grid of models containing the three gases in a wide range of relative molecular abundances. 

Our analysis shows that a model with VMR(CO) = VMR(H$_2$O) = 10$^{-4.5}$, VMR(\methane) = 10$^{-9.5}$, and a steep lapse rate of d$T$/d$\log(p) \sim 330$ K per pressure decade is the best fit to the atmosphere of HD~179949~b, meaning that it gives the strongest cross-correlation signal, corresponding to an S/N~=~6.3. Statistical tests on the measured signal and on the properties of the cross-correlation noise are presented in Sect.~\ref{hd179_stats}, and lead to a significance of 5.8$\sigma$, in line with the above estimate.
The S/N of the detection is only weakly dependent on the atmospheric lapse rate, meaning that it decreases by only $\sim$0.5 when reducing the lapse rate by a factor of 3.

\begin{figure}[h]
\centering
\includegraphics[width=8cm, height=6cm]{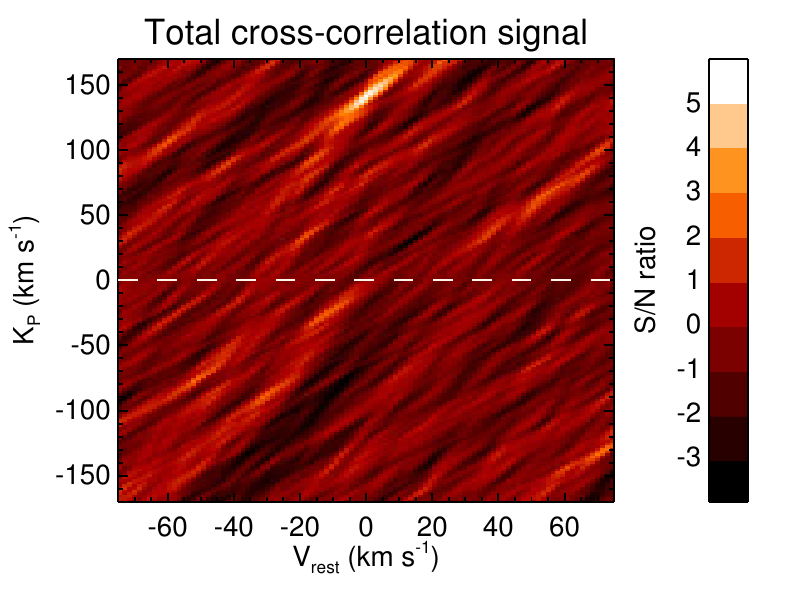}
\caption{Total cross correlation signal from our best-fitting model (see Sect.~\ref{res-three}) for the atmosphere of HD~179949~b, shown as a function of rest-frame velocity and planet projected orbital velocity. The model contains CO and \water\ with equal molecular abundance. The planet is detected with a S/N ratio of 6.3, at a planet projected orbital velocity of \kpvalue, which translates into a planet mass of \mpvalue\ and an orbital inclination of \ivalue.}
\label{all-mols}
\end{figure}

In Fig.~\ref{all-mols} we present the total cross correlation signal from the best fitting model indicated above, which peaks at a planet projected orbital velocity of \kpvalue. The planet signal is retrieved at zero rest-frame velocity when applying a phase shift of $\Delta\varphi=-0.00477\pm0.00069$, which translates into a refined time of inferior conjunction of $T_0=(2,455,757.8190\pm0.0022)$~HJD. Since the stellar RV semi-amplitude is known from previous measurements to be \kstar\ = ($0.1126 \pm 0.0018$) \kms\ \citep{but06}, we can now treat the system HD 179949 as a spectroscopic binary. The mass ratio between the star and the planet is \mstar/\mplan = \kp/\kstar = \mratio, which translates into a planet mass of \mpvalue\ by using the known stellar mass (see Sect.~\ref{sys-description}). Through Kepler's Third Law, we can compute a planet orbital velocity of \vovalue. The orbital inclination of the system is therefore i = $\sin^{-1}$(\kp/$V_\mathrm{orb}$) = \ivalue.

\subsection{Statistical tests on the measured signal}\label{hd179_stats}

The computation of the total cross-correlation signal in Fig.~\ref{all-mols} is extended to negative planet orbital velocities (\kp~$<0$~\kms) to test the robustness of the detection. For negative velocities, which imply an unrealistic retrograde planet orbit, the interaction between the model spectrum and any possible red noise in the data would be similar to the corresponding positive \kp, so false-positive signal would show up as a mirrored image of the \kp~$>0$ plane, while real planet signals would not. In our case, there is no significant signal in the bottom half of the diagram, reinforcing the robustness of the detection.

\begin{figure}[ht]
\centering
\includegraphics{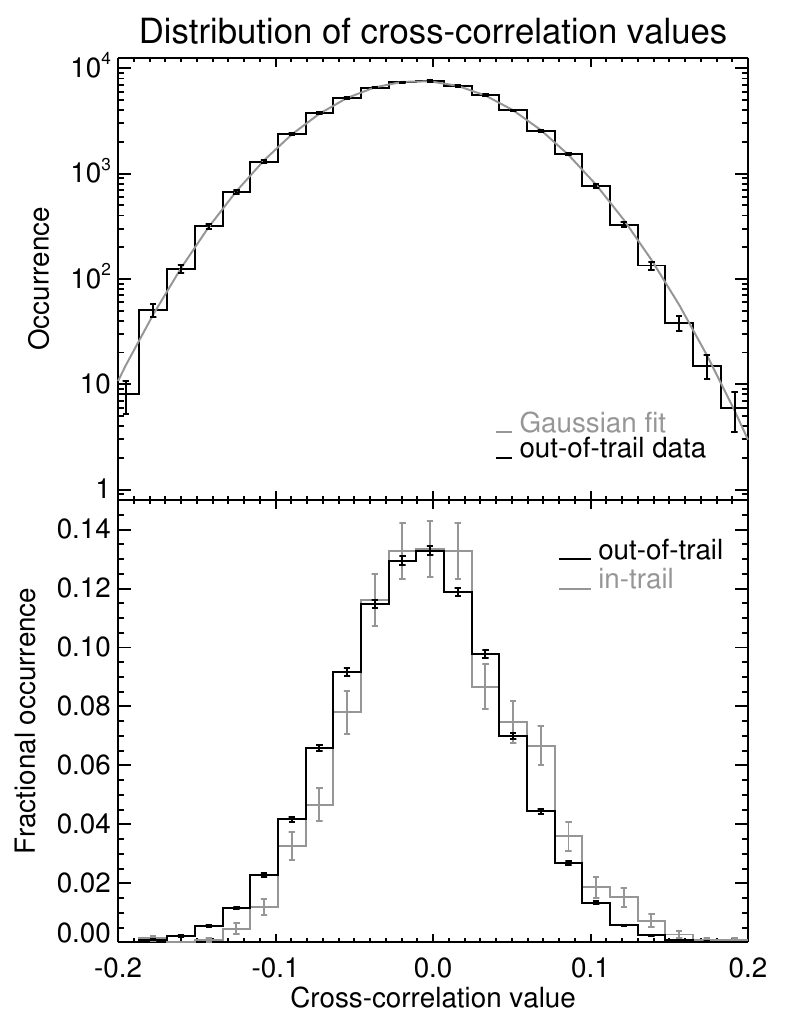}
\caption{{\sl Top panel:} distribution of the cross-correlation values not belonging to the planet RV curve (black histogram) compared to a Gaussian fit (gray line). The two curves are in excellent agreement across the entire range. {\sl Bottom panel:} comparison between the distribution of the cross-correlation values outside and inside the planet RV trail (black and gray histograms, respectively). The latter are systematically shifted toward higher values. A Welch t-test on the data rejects the hypothesis that the two distributions are drawn from the same parent distribution at the 5.8$\sigma$ of confidence level.}
\label{hd179_fig_ccnoise}
\end{figure}

Furthermore, we determined the statistical significance of the signal as in previous work \citep{bro12,bro13,kok13,bir13}. Firstly, we studied the properties of the cross-correlation noise. From the matrix containing the cross correlation signal as function of planet radial velocity and time, CCF($V,t$), we selected those values not belonging to the planet RV curve. In Fig.~\ref{hd179_fig_ccnoise} (top panel) we show their histogram with a black solid line, and a Gaussian fit with a gray line. It shows that the distribution of the cross-correlation noise is in excellent agreement with a Gaussian distribution. Subsequently, we compared the distributions of the cross-correlation values inside and outside the planet RV curve, shown by the black and gray histograms in the bottom panel of Fig.~\ref{hd179_fig_ccnoise}. Even at a visual inspection, it is possible to assess that the distribution containing the planet signal is systematically shifted toward higher values than the distribution of the noise. This can be tested statistically by using a Welch t-test \citep{wel47}. For the signal from HD~179949~b, we were able to reject the hypothesis that the two distributions are drawn from the same parent distribution at the 5.8$\sigma$, in line with the measured S/N.

\subsection{Constraints on the C/O ratio}\label{c-o_ratio}
\begin{figure}[h]
\centering
\includegraphics{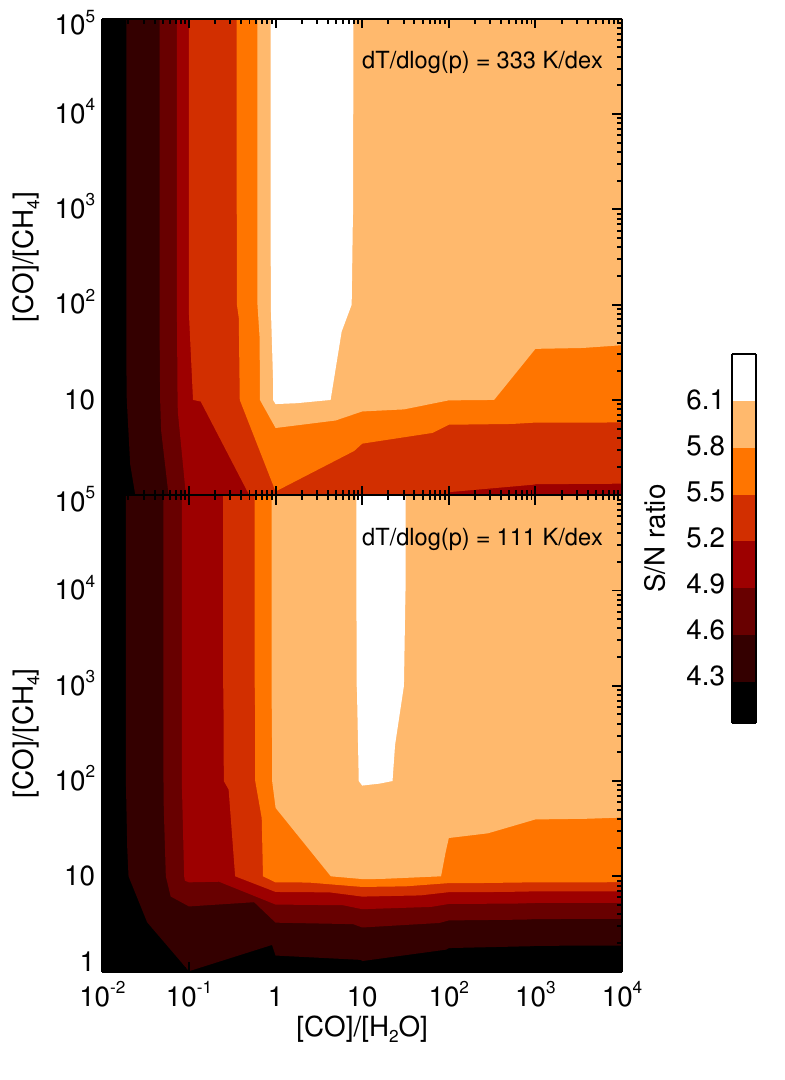}
\caption{S/N of the cross-correlation signal as function of VMR(CO)/VMR(\water) on the horizontal axis, and VMR(CO)/VMR(\methane) on the vertical axis. The results are shown for a lapse rate of 111 K per decade in pressure (lower panel) and 333 K per decade in pressure (upper panel). These are the minimum and maximum values investigated with the atmospheric models containing CO, \water, and \methane.}
\label{sig-vmr}
\end{figure}

In Fig.~\ref{sig-vmr} we plot the measured S/N of the cross-correlation signal as a function of the relative molecular abundances of CO and \water\ on the horizontal axis, and CO and \methane\ on the vertical axis. The top and bottom panels correspond to a steep and shallow atmospheric lapse rate, respectively. As already noticed in Sect.~\ref{res-single}, the strength of the total cross correlation signal is only weakly dependent on the atmospheric lapse rate, so that the top and the bottom plots are qualitatively very similar. Moreover, when CO is highly overabundant over \water\ and \methane, the measured signal to noise approaches that of models with CO alone, i.e., S/N~=~5.9 (upper right quadrants of the two panels in Fig.~\ref{sig-vmr}).

By assuming that CO, \water, and \methane\ are the dominant carbon- and oxygen-bearing molecules at our observed wavelengths, we computed the C/O ratio corresponding to each of our planet spectra as
\begin{equation}\label{eq-co-ratio}
\mathrm{C/O} = \frac{\mathrm{VMR(CO) + VMR(CH_4)}}{\mathrm{VMR(CO) + VMR(H_2O)}}.
\end{equation}
We note that the above assumption neglects any possible contribution from CO$_2$, C$_2$H$_2$, and HCN. Firstly, these molecules cannot be detected with these data, because they are not significant sources of opacity at 2.3~\micron. Moreover, their equilibrium abundances are at least one order of magnitude lower than \methane\ \citep{madhu12}, except for high C/O ratio. This would also imply a high methane abundance, which is inconsistent with our non detection of \methane\ in this data.

In Sect.~\ref{res-three}, we verified that the measured S/N closely matches the significance of the detection as measured by using a statistical approach. Therefore, we can determine 1-$\sigma$ limits on the relative molecular abundances by using the values of S/N, and we can constrain from those the C/O ratio.

We computed the difference between the S/N of each model and that of the model giving the highest cross-correlation, which we denote here with $\Delta_\mathrm{meas}$. In addition, we also computed the expected cross-correlation signal if the planet spectrum would be identical to each of our model spectra. In order to do this, we subtracted the best-fitting model after scaling it in the same way as for measuring the average line depth (see Sect.~\ref{degeneracies}). We then scaled each of the planet spectra to the measured average line depth, and summed it to the data. After running the data analysis for telluric removal, we cross-correlated with the same model, which gave us the expected planet signal. Finally, we computed the difference between the expected signal and the signal measured by cross-correlating with real data, denoted with $\Delta_\mathrm{exp}$.

\begin{figure}[!ht]
\centering
\includegraphics{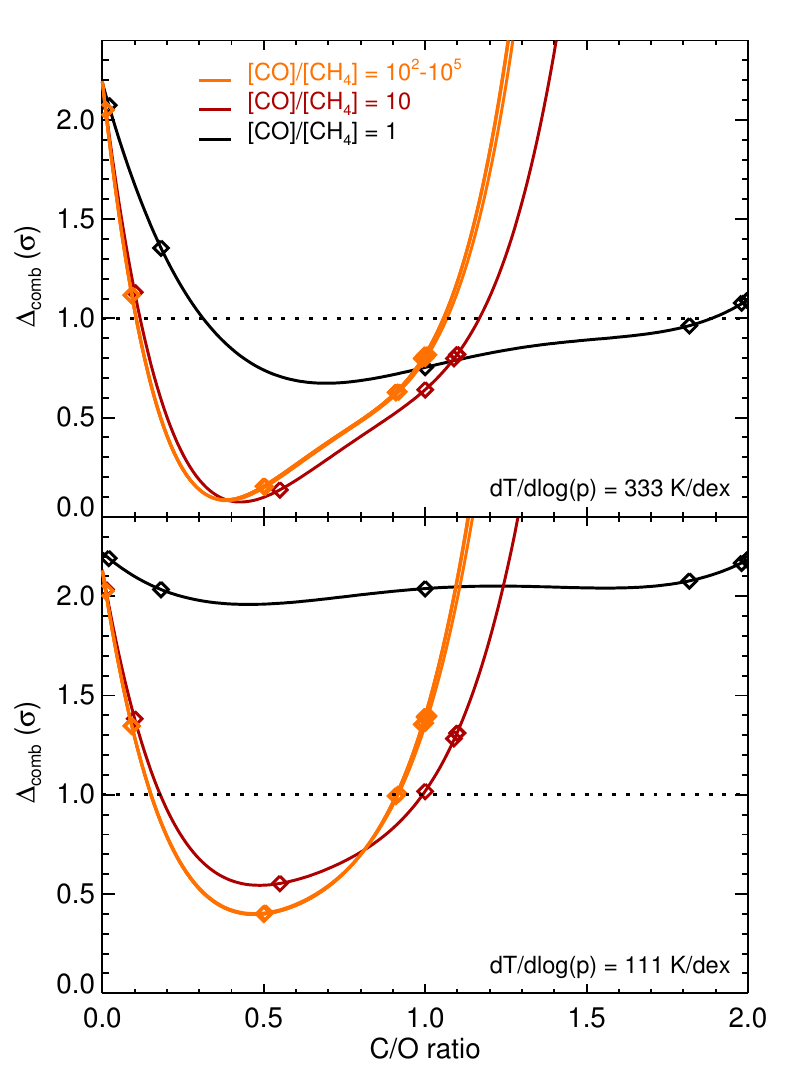}
\caption{Deviation from the best-fitting model ($\Delta_\mathrm{comb}$) as a function of C/O ratio, for atmospheric lapse rates of 333 K/dex (top panel) and 111 K/dex (bottom panel). The C/O ratio is derived from the molecular abundances according to Eq.~\ref{eq-co-ratio}. Since multiple combination of molecular abundances give approximately the same C/O ratio, here we plot curves ranging from a high methane abundance ([CO]/[\methane]~=~1, black curve) to low methane abundance ([CO]/[\methane]~$\ge10^2$, orange curve).  The points are fit with fourth-order polynomials. Except for steep lapse rates with highly-abundant methane, our analysis suggests an oxygen-rich atmosphere for HD~179949~b.}
\label{co-rat}
\end{figure}

By assuming that $\Delta_\mathrm{meas}$ and $\Delta_\mathrm{exp}$ are independent quantities and that they are good proxies for the statistical significance of the signal, we described their combination as a bivariate normal distribution, and we computed the deviation of each model from the best-fitting model according to
\begin{equation}
\Delta_\mathrm{comb} = \sqrt{\Delta^2_\mathrm{meas}+\Delta^2_\mathrm{exp}}.
\end{equation}
In Fig.~\ref{co-rat} we plot $\Delta_\mathrm{comb}$ as a function of the planet C/O ratio, for atmospheric lapse rates of 333 K/dex (top panel) and 111 K/dex (bottom panel). The region above $\Delta_\mathrm{comb} = 1$ denotes the models which deviate from the best-fitting model by more than 1-$\sigma$. 
Except for the case of highly-abundant methane ([CO]/[\methane]~=~1, black line in Fig.~\ref{co-rat}) and a steep lapse rate, our analysis favors a C/O~$<1$, with a best fitting value of C/O$=(0.5^{+0.6}_{-0.4})$. Our analysis therefore suggests an oxygen-rich atmosphere for HD~179949~b.

We note that the non-detection of \methane\ loosens any constraint on the C/O ratio based on the above analysis. This is because the cross-correlation is almost insensitive to progressively increasing abundances of methane, unless its volume mixing ratio equals that of CO. This also means that our estimate of the C/O ratio could potentially be biased due to inaccuracies in the high-temperature methane line list. These could wash out the cross-correlation signal from \methane, and/or increase its detection threshold.

\section{Discussion and future prospects}\label{discussion}

Thanks to high-resolution spectroscopic observations in the near-infrared, we were able to detect molecular absorption from carbon monoxide and water vapor in the atmosphere of the non-transiting exoplanet HD 179949 b. By tracing the planet radial velocity directly, we also solved for the planet mass and the system inclination. 

We do not observe a significant signal when cross-correlating with models containing a thermal inversion. This suggests that overall the temperature decreases with altitude in the observed range of the dayside planet atmosphere. Although weak thermal inversion at high altitude is still possible, these observations exclude the presence of a strong thermal inversion, which would produce molecular lines observed in emission. 
We note that our assumption of an inversion layer extending down to pressures of 1 bar might be too simplistic and not representative of realistic temperature inversions. In particular, the ability of observing emission lines in the spectrum is dependent on the typical atmospheric pressures probed by the CO template as a whole. In order to exclude that our measured lack of thermal inversion is a product of the particular parametrization chosen for the planet atmosphere, we computed a set of models based on the best-fitting inverted $T/p$ profile presented in \citet{hd179_burr07} for the exoplanet HD~209458~b (see their Fig.2, green line), which is often considered the prototype of planets with atmospheric inversion layers. For these calculations we assumed VMR(CO)~=~(10$^{-5},10^{-4},10^{3}$). In line with the previous analysis, we do not detect any planet signal by cross-correlating with these models. 

\begin{figure}[t]
\centering
\includegraphics{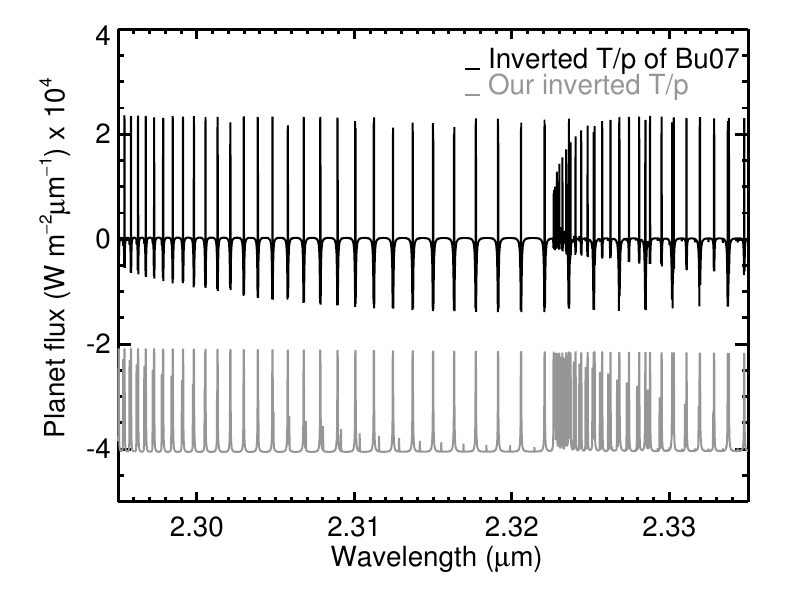}
\caption{Comparison between the best fitting $T/p$ profile of \citet{hd179_burr07} for the atmosphere of HD~209458~b (top curve) and our models with temperature inversion, for a VMR(CO)~=~10$^{-4}$. In both cases, emission lines due to the thermal inversion layer are clearly visible and dominate the narrow component of the planet spectrum.}
\label{hd179_burrows}
\end{figure}

\begin{figure}[h]
\centering
\includegraphics{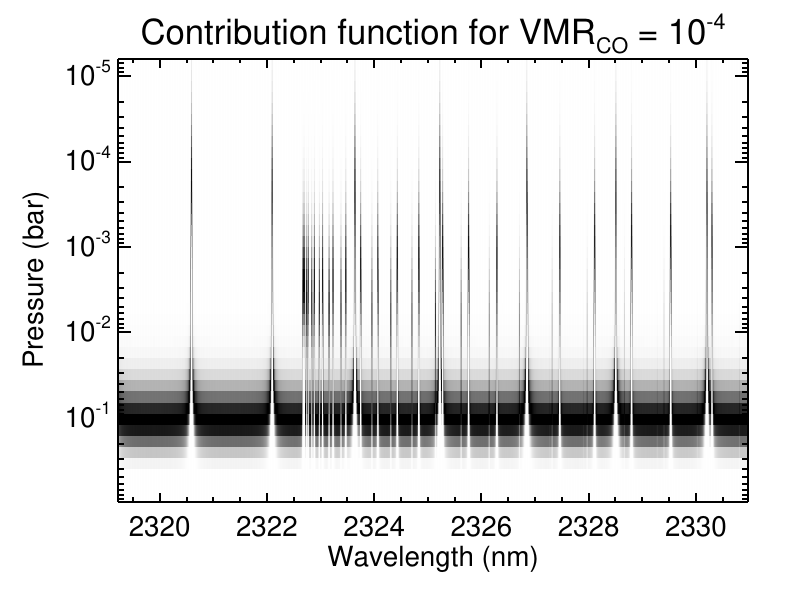}
\caption{Contribution function for the model spectrum of HD~179949~b computed with the \citet{hd179_burr07} $T/p$ profile (Fig.~\ref{hd179_burrows}, top curve). The color scheme runs from zero contribution (white) to maximum contribution (black). The broad component of the spectrum is formed at relatively high pressures (deeper in the atmosphere) and typically probes a non-inverted portion of the $T/p$ profile. On the other hand, the narrow component of the spectrum, to which we are sensitive at high resolution, probes several orders of magnitude in pressure and encompasses thermal inversion layers as well, producing molecular lines in emission.}
\label{hd179_cf}
\end{figure}

Fig.~\ref{hd179_burrows} shows one of these models (for a VMR of CO of 10$^{-4}$) compared to the corresponding model spectrum utilized in this paper. The only qualitative difference produced by a more realistic $T/p$ profile is that the line wings are seen in absorption, while the line core shows emission. Overall the template still shows strong emission lines, and this happens because at very high resolution the narrow component of the planet spectrum (which is the component detected through cross-correlation) probes pressures several order of magnitude lower than those of the continuum part, including the typical altitudes at which thermal inversion layers are thought to form. This is illustrated by Fig.~\ref{hd179_cf}, showing the CO contribution function for the model based on the \citet{hd179_burr07} profile. Although the line wings probe deep in the atmosphere, where realistic thermal inversions might not persist, the sensitivity to this part of the spectrum is very low due to our data analysis (see Sect.~\ref{degeneracies}). We instead remain sensitive to the contrast between the broad and the narrow component of the spectrum, which probes 4-5 orders of magnitude in pressures as a whole and encompasses the thermal inversion layers as well, corresponding to molecular lines observed in emission. We can therefore conclude that, based on our analysis, HD~179949~b does not show a HD208458b-like inversion.

Thermal inversion layers might be influenced by the incident stellar flux \citep{for08}, with the most irradiated planets being capable of retaining in the gas phase high-altitude optical absorbers, such as TiO or VO, whereas these compounds would condense and rain out at lower irradiation levels. The planet HD 179949 b receives a stellar flux of $\sim 1.4\times 10^9$ erg cm$^{-2}$ s$^{-1}$, which according to the \citet{for08} classification should be enough to keep optical absorbers in the gas phase. However, stellar activity could also influence the onset of thermal inversion layers, because of the UV stellar flux potentially destroying the above-mentioned absorbers \citep{knu10}. With an activity index of $R'_{HK} = -4.72$, HD 179949 can be considered an active star, so that the observed lack of a thermal inversion could fit in this scenario. However, it has recently been pointed out that a potential key factor in determining the equilibrium chemical composition of a hot Jupiter is the C/O ratio \citep{madhu12}. For carbon-rich planets, there is little oxygen left for the formation of oxides, meaning that molecules like TiO and VO would be under-abundant in these atmospheres, and could not cause a thermal inversion layer. 

The observational evidence for a carbon-rich planet has been presented for the hot-Jupiter WASP-12~b, based on photometric measurements of its secondary-eclipse depth \citep{madhu11}. Subsequently, the discovery of a faint nearby star diluting the light of the system revised the slope and the shape of the dayside spectrum, so that the assumption of a high C/O ratio, although not inconsistent with the data, is not the only possible explanation for the planet atmosphere \citep{cro12}.

In this work we derive limits on the relative molecular abundances of CO, H$_2$O, and CH$_4$, and we translate them into a measurement of the atmospheric C/O ratio. However, due to the fact that around 2.3~\micron\ CO dominates the cross-correlation signal and that CH$_4$ is not detected, only loose limits on relative abundances could be derived. These are insufficient to robustly constrain the C/O ratio, so that the hypothesis of a carbon-rich planet as explanation for the lack of a thermal inversion layer cannot be tested yet, although our analysis tends toward a low C/O ratio.

In order to improve these measurement, one would need to detect multiple molecules at multiple wavelengths. For example, as shown in \citep{remco14}, the region around 3.5 \micron\ is expected to contain the detectable signatures of H$_2$O, CO$_2$, and CH$_4$. By linking observations at 2.3~\micron\ and 3.5~\micron, it is in principle possible to measure the relative abundances of the four major C- and O-bearing molecular species in hot-Jupiter atmospheres, which would definitely improve constraints on the planet C/O ratio.

Moreover, for molecular species such as water vapor and methane, which absorb across a wide spectral range, high-resolution spectrographs capable of covering one or multiple NIR bands in a single setting would be beneficial to these observations, since the cross-correlation signal scales roughly as the square root of the number of strong molecular lines. The upgrade of CRIRES to a cross-dispersed spectrograph will increase the spectral range of the instrument by a factor of $\sim$6. Moreover, although mounted on smaller telescopes, more NIR spectrographs with an even wider spectral range, such as CARMENES \citep{quir12} and SPIRou \citep{spirou12}, will come on-line in the northern hemisphere.

\begin{acknowledgements}
We thank Simon Albrecht and Ernst de Mooij for their contribution to the design of high-resolution observations in the near-infrared, and for the constructive discussion about the results from our large program. We thank Sergey Yurchenko and Jonathan Tennyson for their useful insights on the comparison between the EXOMOL and HITRAN spectroscopic line list of methane. R.~d.~K acknowledges support from the Netherlands Organisation for Scientific Research (NWO).
\end{acknowledgements}

\bibliographystyle{aa}
\bibliography{refs}

\end{document}